# Liberating Logic in the Age of AI:
# Going Beyond Programming with Computational Thinking

Douglas C. Schmidt and Dan Runfola
William & Mary
Williamsburg, VA, USA

**Abstract**

*Mastering one or more programming languages has historically been the gateway to implementing ideas on a computer. Today, that gateway is widening with advances in large language models (LLMs) and artificial intelligence (AI)-powered coding assistants. What matters is no longer just fluency in traditional programming languages but the ability to think computationally by translating problems into forms that can be solved with computing tools. The capabilities enabled by these AI-augmented tools are rapidly leading to the commoditization of computational thinking, such that anyone who can articulate a problem in natural language can potentially harness computing power via AI.*

*This shift is poised to radically influence how we teach computer science and data science in the United States and around the world. Educators and industry leaders are grappling with how to adapt: What should students learn when the hottest new programming language is English? How do we prepare a generation of computational thinkers who need not code every algorithm manually, but must still think critically, design solutions, and verify AI-augmented results?*

*This paper explores these questions, examining the impact of natural language programming on software development, the emerging distinction between programmers and prompt-crafting problem solvers, the reforms needed in computer science and data science curricula, and the importance of maintaining our fundamental computational science principles in an AI-augmented future. Along the way, we compare approaches and share best practices for embracing this new paradigm in computing education.*

## 1. Introduction

**The coming commoditization of computational thinking**. Interest in computing careers has historically risen and fallen in distinct waves (Microsoft Research 2006). In the early 1980s, a surge in computer science (CS) enrollment was fueled by job-market anxiety ("If I don't have a CS degree, I won't get a job"). This assumption soon proved unfounded, leading to a sharp decline in majors by the late 1980s. By the late 1990s, amid the dot-com boom, students again gravitated toward CS, this time motivated by visions of entrepreneurial wealth ("If I don't major in CS, I won't be a millionaire!"). When the bubble burst, CS enrollments declined sharply once more.

Beginning around 2015, a diverse ecosystem of programming-centric educational pathways emerged alongside resurgent interest in computing driven by dramatically lower costs and increased accessibility (e.g., due to the emergence of simple-but-powerful scripting languages). Dedicated data science (DS) programs were one visible example, but they were accompanied by domain-specific computing tracks in fields like biology, engineering, the social sciences, and the humanities, as well as shorter certificate and bootcamp offerings. These initiatives started modestly but expanded steadily, sometimes attracting a higher proportion of women in their early cohorts.

At the same time, CS enrollments rose again, buoyed in part by this broader surge in the use of computation as a tool for inquiry. Today, another structural shift is underway: generative AI is beginning to change the substance of coding, creating new pressures and opportunities for CS and DS degree programs to rethink their curricula and outcomes.

This structural shift is visible in AI assistants that generate code and analyze information, blurring the line between program writing and program using. For example, research tasks in many disciplines can be delegated to AI agents that autonomously browse sources, explore datasets, and synthesize answers, thereby acting like tireless virtual post-docs. Non-experts can rapidly generate non-trivial software programs, including web user interfaces and common data modeling and visualization solutions (O'Reilly 2025). These advances compel reexamining long-standing assumptions in computing – how we define it, teach it, and use it.

This paper re-examines computing's future, addressing questions like how should the field evolve when students can solve problems in natural language? What becomes of developers when computational thinking itself is commoditized? And which emerging skills—like prompt engineering (White et al. 2023) and prompt-ware engineering (Chen et al. 2025)—will define success? We explore these questions by tracing the rise of prompt-driven application development, shifts in professional roles, curricular implications, and the risks of over-relying on AI-augmented tools.

**Paper organization**: The remainder of this paper is organized as follows: Section 2 examines natural language as a new programming paradigm, enabling problem-solving through prompts and commoditizing computational thinking; Section 3 explores how AI is reshaping knowledge work, rewarding those who combine critical thinking and domain expertise with AI fluency; Section 4 outlines the maturation of prompt engineering into a disciplined practice with testing, version control, and maintainability; Section 5 discusses evolving CS and DS curricula that merge core computing principles with AI-era skills, positioning computational thinking as universal literacy; Section 6 warns that human judgment remains essential to mitigate risks, uphold ethics, and sustain authentic problem-solving; and Section 7 concludes by emphasizing that success in AI-augmented computational thinking depends on fusing human creativity with machine intelligence to address challenges ethically and effectively.

## 2. From Coding to Natural Language: The Evolution of Computational Thinking

Jeannette Wing (Wing 2006) popularized the term "computational thinking" to describe a problem-solving approach grounded in logic, abstraction, decomposition, pattern recognition, generalization, and automation. Historically, applying computational thinking required learning to program, by translating ideas into code using languages like Python, Java, or C++. Despite the popularity and impact of this approach to problem solving, the requirement to "learn how to code" has long presented a daunting hurdle for many people.

For example, mastery of programming languages meant grappling with myriad incidental complexities, including arcane syntax and semantics, debugging errors, memory management, and other low-level issues unrelated to the core domain problem. These barriers have long constrained who could transform ideas into functioning software applications. They have also limited how broadly computational thinking could be taught, since students were required to master the intricacies of programming alongside the conceptual practices of the field.

A useful analogy comes from the evolution of automobiles. Over a century ago, driving required knowing how to build a car; decades later, it meant understanding how to maintain one. Today, it typically involves little more than steering. Moreover, even steering is becoming optional as automobiles rapidly approach full autonomy.

Computing has followed a similar arc. In its early days, users often had to assemble hardware or write low-level code to accomplish basic tasks. As recently as a few decades ago, programming largely involved writing and debugging code manually in third-generation languages like Python, C++, or Java. Over time, high-level libraries, frameworks, and tools have alleviated much of that burden, just as automatic transmissions and power steering simplified driving. The latest leap, however, is far more transformational: *natural language is emerging as a viable programming abstraction*. Today, users can describe desired outcomes in English or other human languages and AI systems can generate and execute corresponding solutions.

The paradigm of translating a natural-language problem description into a working solution without traditional coding is already a reality (Michaelsen/Santos 2024). For many tasks, users now achieve results by crafting prompts instead of hand-coding algorithms. Researchers and professionals in non-computing domains who once had to learn programming to analyze data or build tools can now leverage conversational AI assistants instead. For example, a biologist or sociologist can pose questions to LLMs and receive data analyses, visualizations, or simulations without writing a single line of Python or R.

Figure 1 shows how the barrier-to-entry inversion enabled by LLMs commoditizes computational thinking by abstracting away programming expertise, putting powerful tools into the hands of non-programmers. This figure captures the shift in who can engage meaningfully in computational thinking tasks like a big-data statistical analysis before and after the emergence of LLMs (Cerf 2023). In the "Before LLMs" era on the left of the figure, the barriers to entry were steep: only computer scientists and data scientists had relatively easy access to such capabilities, while domain experts and the general public faced high technical thresholds. In the "After LLMs" era on the right of the figure, however, the barrier to entry is lowered across the board, especially for domain experts who already have the requisite knowledge to pose meaningful questions and can now perform once-esoteric tasks by submitting natural language prompts to modern LLMs.

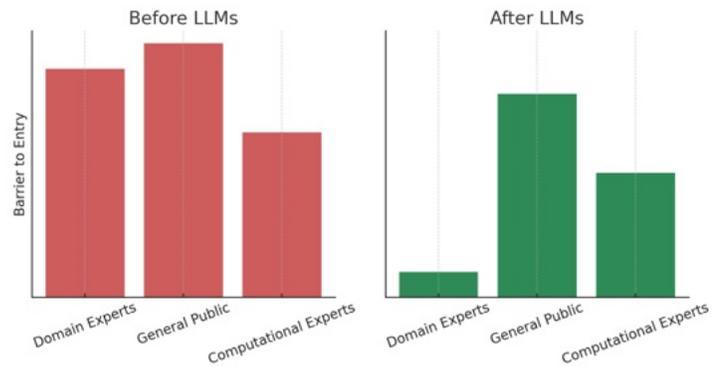

*Figure 1: Barrier-to-Entry Inversion: An Example of the Difficulty of Big-Mata Modeling Before and After LLMs.*

Figure 2 visualizes another aspect of this commoditization: by collapsing the stack from problem formulation to output production, LLMs make it possible for anyone with a question, some curiosity, and an internet connection to access the powers once reserved for those with significantly more technical expertise. There are an emerging collection of examples in the literature representing this approach, including automated generation of visualizations (Vázquez 2024, Khan 2025, Yang 2024), cluster biological data (Longyi 2025), simulate economic scenarios (Nian 2024), and optimization techniques (Schmidt 2025a). With prompt refinements, tools like ChatGPT can produce satisfactory results in a fraction of the time needed to implement such tasks manually in Python or R. This productivity acceleration underscores how dramatically the barrier to entry for computational problem-solving has been lowered. Domains that once required custom programming for analysis or simulation can now benefit from this new mode of computational thinking. Computational thinking is thus no longer scarce, it's everywhere. Some call this its democratization, others its commoditization, but either way the proliferation is undeniable: generative AI has made it a baseline skill for everyone, not just coders.

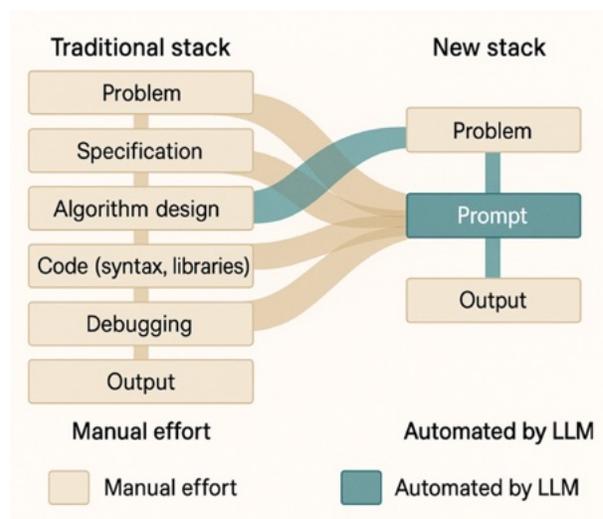

*Figure 2: Collapsing the Computational Thinking Stack.*

3. **Knowledge Work at a Crossroads: AI and the Job Market**

The commoditization of computational thinking is already reshaping the job market. Tasks once associated with knowledge work, such as analysis, coding, and writing, are increasingly automatable at scale. This trend marks a historic inflection point, as shown in Figure 3. Earlier waves of automation displaced factory labor and service work; today, AI is encroaching on roles long considered secure behind a desk.

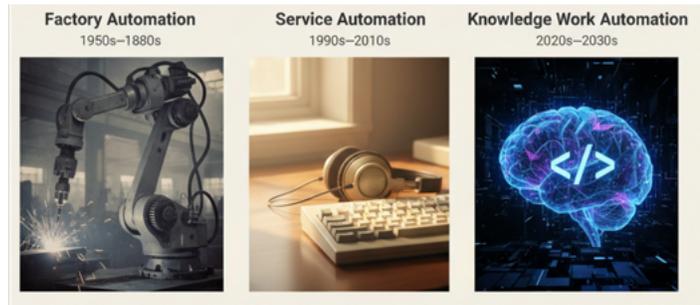

*Figure 3: The Expanding Scope of Automation.*

Many knowledge workers are understandably concerned about what AI portends for their careers. The familiar adage holds: "AI won't take your job - but someone who knows how to use AI might." In other words, those who skillfully leverage AI tools can dramatically outperform peers who do not.

For instance, a financial analyst using AI to generate visualizations and draft reports can accomplish in hours what would traditionally take days. Similarly, a programmer working with an AI code-assistant can deliver working, robust features far more rapidly than one coding everything manually. By contrast, those who lack AI-augmented workflows increasingly struggle to match the productivity *and* quality of their AI-augmented peers.

As shown in Figure 4, the distribution of developer effort has shifted markedly in the era of generative AI. Empirical studies prior to the advent of AI-augmented coding assistants indicated developers devoted up to 70% of their time to manual coding, testing, and debugging, with the remaining 30% focused on framing and understanding the problem (Stripe/Harris 2018; Kumar et al. 2024).Today, those proportions are being rebalanced: manual coding now occupies roughly 20% of a developer's effort, while AI prompting, oversight, and orchestration account for approximately 50% (Mozannar et al. 2024; Software.com 2023). Notably, the share of time spent on problem framing remains at about 30%, but its strategic importance has grown as developers increasingly guide, evaluate, and integrate AI-generated solutions more (GitHub 2024). In this new landscape, the developer role evolves from producing code to managing and validating an AI-driven workflow.

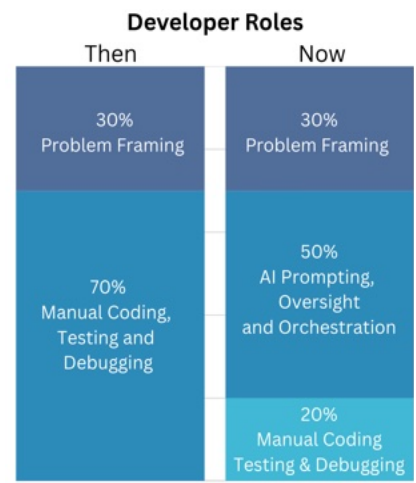

*Figure 4: Redefining Developer Roles.*

In short, natural language has become the new compiler, and developer's focus is migrating from syntax and semantics to strategy. This transformation raises a provocative question: if working solutions can be produced via strategic dialogue with an AI assistant, does that not also qualify as programming? The implication is that computational thinkers, i.e., those adept at framing problems and directing AI, may soon outnumber—and in some domains outperform—traditional software engineers in the workforce, even without traditional programming instruction.

In a workforce paradigm dominated by computational thinkers, humans increasingly serve as the architects, guides, and editors of AI-produced work and *not* as traditional coders. The process begins with humans defining the problem in context, AI producing an initial draft—whether code, analysis, or other products—and humans then refining and validating the result. Creative ideation, ethical considerations, and architectural decision-making remain distinctly human-led, while repetitive implementation and routine knowledge retrieval are increasingly delegated to AI. This shift is already redefining how both students and working professionals view their identities and career trajectories in technology.

For example, a recent article (Fore 2025) observed that "Nearly half of Gen Z and millennials say college was a waste of money. AI has already made degrees obsolete." This trend presents a dilemma for aspiring tech workers: pursue a traditional CS degree or take the route of learning AI skills through other paths (e.g., self-directed learning with AI tools). The traditional CS degree route delivers strong fundamentals and deep conceptual grounding, but it comes with high costs, a time commitment of roughly four years, and a slower start in the workforce. In contrast, AI-assisted, self-directed learning paths offers rapid entry and immediate practical fluency, but risks leaving significant gaps in the underlying theory and problem-solving depth that seasoned engineers rely on.

This shift is forcing universities, students, and employers to reassess how they define "career readiness" in the age of generative AI. If foundational education risks being undervalued, those who bypass it may find themselves hitting ceilings when AI tools alone can't solve complex problems. Those best positioned for long-term success may be those who combine the speed and agility of AI-augmented learning with the rigorous thinking and domain mastery traditionally gained through formal study. That blend of fast, adaptive skill acquisition built on a bedrock of core principles will likely define the most resilient and versatile tech professionals of the coming decade.

### 4.   Prompt Engineering: From *Ad Hoc* Queries to a New Discipline

The practice of prompting has advanced through distinct phases over the past few years, as shown in Figure 5.  In the early days of the current AI boom—less than two years ago—prompting was often dismissed as a gimmick, a fleeting "flash in the pan" for extracting useful answers from LLMs (Morris 2024). At that stage, prompting was largely an *art*, i.e., a solo, trial-and-error process guided by intuition rather than disciplined methods. Over time, however, practitioners began to realize that carefully structured natural language inputs could reliably shape model outputs. Prompt engineering came to be understood as programming *through* the AI, rather than programming the AI itself, with wording, context, and sequencing exerting significant influence on the quality and reliability of responses (White et al. 2023).

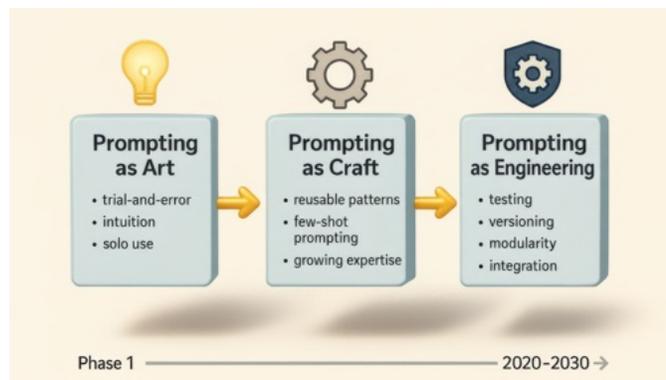

*Figure 5: From Promoting as Art to Prompting as Engineering.*

The middle stage, shown in Figure 5 as *Prompting as Craft*, reflects the maturation of this skill into a repeatable methodology. Seasoned users now leverage reusable prompt patterns, which are tested phrasing techniques that consistently elicit higher-quality or structured outputs (White et al. 2024). Techniques like "chain-of-thought reasoning" prompts (where an LLM is directed to articulate its step-by-step logic before answering), "few-shot" prompting (where examples precede a new request), and "planning" (where explicit sub-goals are articulated by the LLM to reference during inference) allow users to guide LLMs toward more logical, context-aware, consistent and transparent results. At this stage, prompting becomes less about improvisation and more about applying accumulated expertise to achieve predictable and accurate outcomes.

Looking ahead, Figure 5 anticipates a third phase: *Prompting as Engineering* (Chen et al. 2025) or "prompt engineering in-the-large." Here, prompting transcends individual experimentation to become a scalable, professional discipline. This phase involves integrating prompting into larger systems with rigorous practices, such as testing, version control, modularity, and system-wide integration, which mirror the historical evolution of software development from *ad hoc* coding to structured engineering. The difference between crafting a clever one-off prompt and designing a robust, mission-critical AI pipeline is akin to the gap between writing a single script and architecting a complex software-reliant system.

Scaling the discipline of prompt engineering requires adopting many of the disciplined practices that matured in traditional software engineering, as shown in Figure 6. Just as version control, testing, documentation, code reuse, and modular integration became essential to managing complexity in conventional development, prompt engineering in-the-large must embrace analogous principles and require deliberate attention to the following practices:

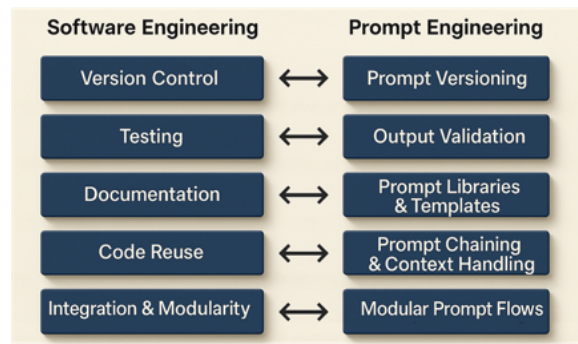

*Figure 6: Software Engineering vs. Prompt Engineering.*

- **Quality assurance**: Verifying that AI outputs are accurate, relevant, and appropriate for their intended context, not merely accepting the first response generated.

- **Testing and validation**: Exploring alternate phrasings and input scenarios to assess consistency and uncover potential failure cases.

- **Maintenance**: Updating prompts as models evolve or task requirements shift and tracking version history, which is akin to maintaining traditional source code.

- **Integration**: Ensuring AI-generated components integrate smoothly with each-other and traditional systems, including error handling, prompt chaining, and context continuity.

These practices are nascent but will become foundational to reliable AI-driven development (Ticong 2024). Without these practices, teams building complex systems through prompting risk accumulating significant technical debt (Menshawy 2024). Early adopters note that *ad hoc* prompt-hacking often produces brittle solutions (Morris 2024). Just like traditional software engineering, without documentation, standards, or modularity, even small shifts in model behavior—or changes in team members' practices—can lead to cascading failures.

Conversely, approaching prompt engineering with a scalable engineering mindset unlocks significant potential. For instance, teams can curate libraries of validated prompt templates—analogous to software libraries—for recurring tasks like data extraction, code generation, or style transformation. Large-scale projects may institute prompt design reviews akin to code reviews or even unit tests to ensure AI components are reliable, interpretable, and aligned with fairness goals.

Over time, we expect AI tooling to mature, along with emerging patterns for orchestrating multiple AI agents within a single application. Integrated development environments (IDEs) are already offering features to debug, trace, and optimize prompt flows. In short, prompt engineering is evolving from *ad hoc* experimentation into a rigorous discipline poised to become central to future software-reliant application and system development.

## 5. Rethinking Education for an AI-Augmented Era

Few domains feel the impact of commoditized computational thinking more acutely than higher education. Faculty across the computational sciences now face the challenge of preparing students for a world where conventional coding is no longer the primary means of, or barrier to, implementation. This challenge raises a foundational question: Should everyone still learn to code? For now, the answer remains "yes," but not for the same reasons, nor in the same way as before.

***Impact of AI on computer science education***. In an era of AI copilots and conversational IDEs, traditional software development skills are no longer sufficient. Instead, students must learn to *collaborate* with AI by leveraging tools like Codex, Copilot, Cursor, Zed, or Windsurf to design solutions, craft effective prompts, and critically evaluate the results. CS curricula must therefore expand to teach prompt engineering and AI-augmented problem solving alongside coding with conventional programming languages.

Introductory CS courses have historically (and reasonably) emphasized writing programs from scratch in languages like Java, C++, or Python and debugging them manually, reflecting an era when coding was the primary pathway into computational thinking. Today, however, focusing solely on manual coding risks leaving the curriculum outdated, as shown in Figure 7. Students recognize that many tasks once considered essential can now be performed more quickly and reliably by AI tools and AI-augmented IDEs.

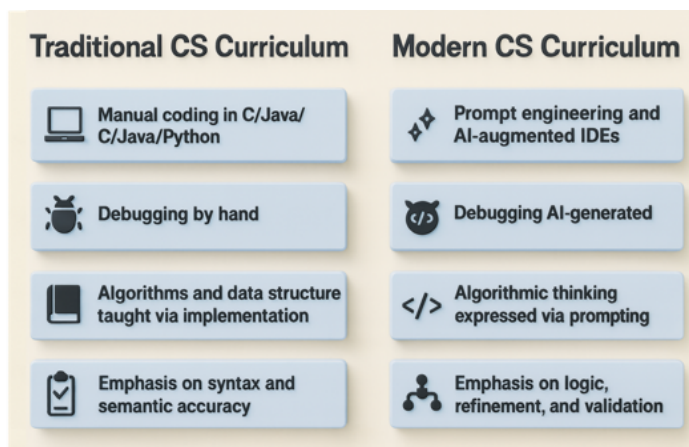

*Figure 7: Traditional vs. Modern CS Curriculum.*

Universities that persist in teaching CS as they did decades ago will struggle to attract and retain students in an era defined by instant, AI-generated solutions. To remain relevant and prepare graduates for the workforce, therefore, educators must integrate AI-driven tools into their pedagogy. Rather than dismiss these technologies as shortcuts or "cheating," forward-looking instructors are framing natural language programming and prompt engineering as core components of modern computing practice (Porter & Zingaro 2024).

For instance, an introductory programming course might incorporate AI-augmented IDEs like Windsurf or Cursor, where students generate code with AI assistance but must still explain, refine, and validate the results. This approach, reflected in the "Modern CS Curriculum" side of Figure 7, shifts the emphasis from the conventional focus on syntax and hand-implementation to problem definition, logic, refinement, and verification. In doing so, students can build the skills necessary to thrive in an AI-augmented workforce where computational thinking increasingly takes the form of design and prompting rather than manual coding (Delaney 2025).

Educators should avoid encouraging blind reliance on AI tools (Denny et al. 2024). Instead, they must develop discerning programmers who use AI to accelerate their work while still ensuring quality (O'Reilly 2025). Effective practice follows an iterative loop shown in Figure 8: framing the problem, prompting the AI, reviewing outputs with domain expertise, and testing for correctness before refining prompts or writing code manually. This mirrors long-standing "computational thinking," but now the challenge is teaching it without the traditional anchor of syntax and programming—an unfamiliar paradigm that demands new instructional approaches.

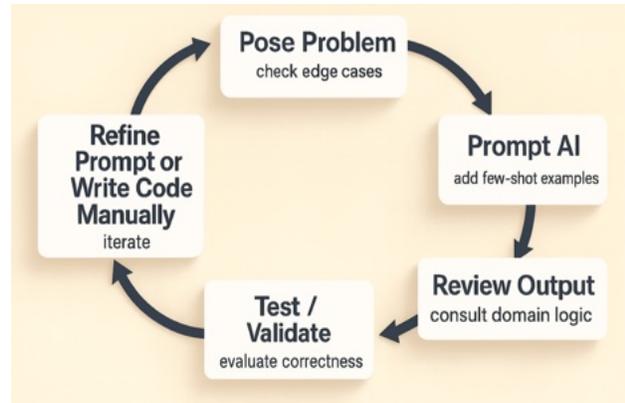

*Figure 8: AI-Augmented Development Requires Iteration.*

Students trained in this iterative "trust but verify" approach that tests AI outputs, checks edge cases, and refines solutions will be much better prepared than those who simply accept an AI tool's initial output. Developing these habits requires *metacognition* to monitor how AI reasons, identify flaws or gaps, and intervene when needed. In essence, computational thinking must be redefined for the AI era. For the time being, graduates often still require robust mental models of algorithms and data structures. Increasingly, however, they will apply those models in dialogue with intelligent systems, rather than solely through manual coding in traditional IDEs.

**Impact of AI on data science education**. Data science is grounded in statistical reasoning and computational methods, so it too faces major disruption (Tu et al. 2024). AI can now automate many traditionally time-inten-sive steps—ETL pipelines, data cleaning, exploratory analysis, and even initial model construction, as shown in Figure 9. Rather than scripting in Pandas or tweaking model syntax, data scientists can issue natural-language instructions and receive draft analyses, visualizations, and models. As a result, DS success will depend less on coding mechanics and more on the ability to frame problems in ways AI systems can operationalize and extend.

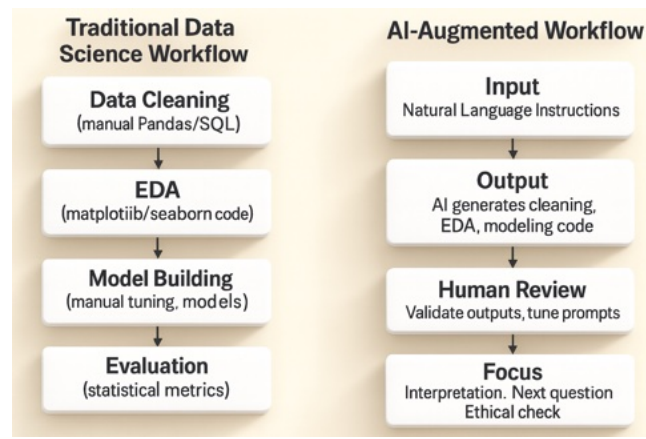

*Figure 9: Yesterday's Data Science vs. Tomorrow's.*

This shift also makes teaching computational thinking central. With routine engineering tasks offloaded to automation, the differentiating skill for new data scientists is the capacity to abstract business or research problems into data-driven forms that interface effectively with large language models, and then to validate and interpret the results. Senior data scientists of tomorrow will need to define what kinds of data are required to solve a given problem, how to collect them at reasonable cost, how to verify that AI-driven pipelines operate as intended, and how to analyze outputs downstream to generate meaningful insights. As highlighted in Figure 9, the discipline is moving from "How do I build the pipeline?" to "How do I think about the problem so the pipeline and the AI behind it produces something I can trust?"

Rather than diminishing the value of fundamentals, the advent of AI in data science makes them even more essential. As shown in Figure 10, AI can generate code, models, and visualizations from natural-language prompts, but humans must still review and refine those outputs using their domain knowledge. For example, students should ask: *Is the model valid? Are the assumptions sound? Is the analysis unbiased, cost-effective, and timely?* These questions rely on core statistical reasoning and operational awareness. By integrating AI-augmented tools into the curriculum, educators can train graduates not merely to accept AI results, but to interrogate, validate, and improve them. In this new workflow, success shifts from hand-coding to curating analyses with judgment, context, and rigor.

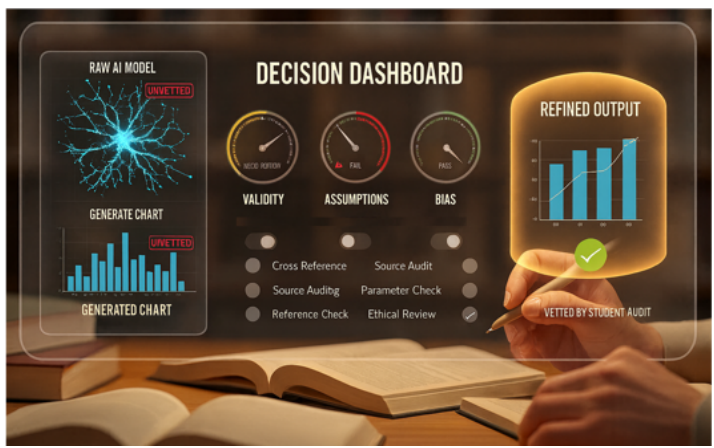

*Figure 10: Humans at the Helm of the AI Pipeline.*

***Rethinking curricula for an AI-augmented era***. Academic institutions are revamping their programs in response to the disruption caused by generative AI, restructuring curricula to both sustain student interest and align with workforce demand. At William & Mary, this shift is already evident: CS has become the fifth most popular major, while DS ranks 13th. Taken together, these programs now surpass every other field in student popularity. Data science is also the most frequently chosen minor - attracting nearly twice as many students as psychology, the next most popular choice - underscoring the growing appetite for computational and AI fluency across disciplines.

As shown in Figure 11, this demand extends well beyond traditional CS pathways. Students in the humanities, social sciences, natural sciences, and professional fields increasingly seek to integrate computational thinking and AI into their studies. Institutions are responding with new interdisciplinary programs that embed AI across a wide range of disciplines. For example, Virginia Commonwealth University (VCU) recently introduced an 18-credit Humanities and Science minor in AI that spans courses in anthropology, English, history, philosophy, physics, statistics, and more, demonstrating how AI literacy is being woven into the fabric of diverse fields of inquiry (VCU 2025).

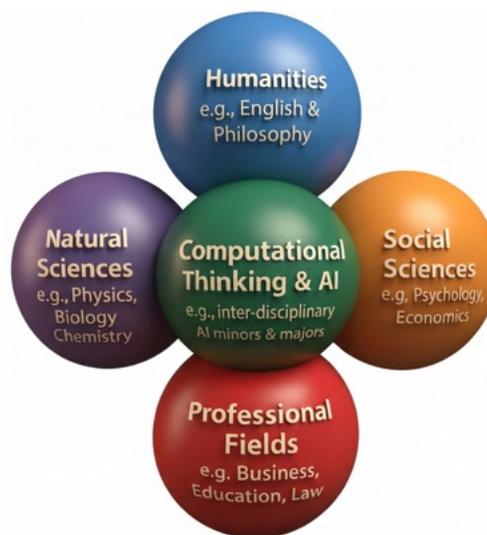

Similar initiatives are underway at William & Mary's School of Computing, Data Sciences & Physics. New majors and minors are being designed that allow students to combine AI fluency and mastery with other degrees. The goal is to make computational thinking and AI skills accessible to students in economics, sociology, psychology, and other disciplines within Arts & Sciences, as well as to those in professional schools. As highlighted in Figure 11, this curricular model recognizes that AI literacy is not just for aspiring technologists, but an essential competency for graduates across every domain of study.

*Figure 11: AI is a Cross-Cutting Competency.*

The three-tier curricular model in Figure 12 shows how computing-related learning is broadening across the university. At the foundation is *AI Literacy*, the widest and fastest-growing layer, encompassing students from every discipline who are learning to read, question, and responsibly interpret AI-generated content. These students—whether in biology, economics, sociology, the arts, or professional fields—may never write code, but they are gaining the essential ability to use AI tools to explore ideas, evaluate evidence, and support domain-specific analysis.

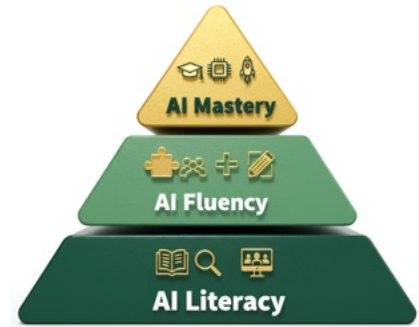

*Figure 12: Hierarchy of AI Proficiency.*

The middle layer, *AI Fluency*, represents students who move beyond literacy to actively shape AI-assisted workflows. These learners blend computational thinking, data skills, and domain expertise. They can design prompts, critique model behavior, integrate analytic tools, and collaborate with AI systems to build and refine solutions. This population includes rapidly growing majors and minors that combine computer science, data science, and discipline-specific computation.

At the apex is *AI Mastery*, a smaller but vital group of students who develop deep technical competence—traditional programmers, computer scientists, and AI specialists who build, optimize, and audit the systems everyone else relies on. They advance the underlying models, ensure reliability and security, and craft novel architectures and tools. Together, these three layers describe a future in which computing education is not a single pipeline but a diversified ecosystem—with AI shaping how every student learns, reasons, and creates.

The implications for education are significant. Universities must design curricula not only for a small group of traditional programmers, but for the far larger population whose competitive edge will come from AI-augmented reasoning. This shift requires teaching students from diverse backgrounds to frame problems for AI, assess machine-generated results, and refine outputs with domain expertise. By embedding AI tools across disciplines and emphasizing adaptability alongside technical fluency, institutions can prepare graduates for a world where computational power is commoditized. In that landscape, the essential skill isn't coding alone—it's the ability to coordinate human judgment with AI capabilities.

## 6. Navigating Limitations: Why Humans Remain Critical

AI-augmented development offers major benefits, but its limitations are real. LLMs and AI tools still carry notable risks as natural-language programming partners (Hüttel 2024), especially oversimplification and hallucination. Early systems often produced confident but fabricated answers when used without careful prompting (Hicks et al. 2024). Even today's models can still confabulate under ambiguity or when pushed beyond their scope—sometimes inventing details or citations that appear credible but are entirely false.

Fortunately, generative AI tools are improving rapidly. Newer LLMs, such as OpenAI's Deep Research, employ multi-step deliberation processes that reduce the likelihood of hallucinations. Yet, as Figure 13 depicts, their outputs remain largely a *black box* that is fast but opaque, lacking the transparency needed for verification.

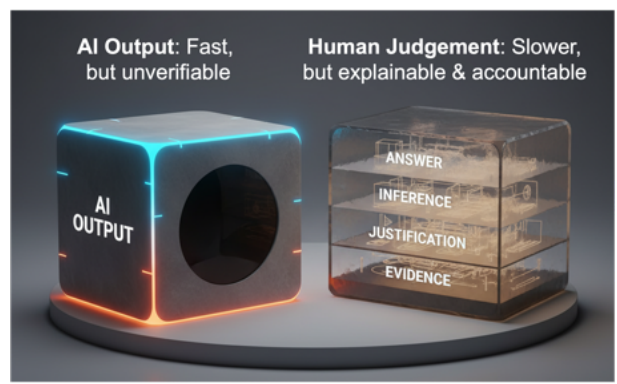

*Figure 13: Speed Without Transparency is Risky.*

On the left side of the figure, the "AI Black Box" produces answers, often without exposing the underlying reasoning, leaving users with little visibility into inference, justification, or evidence. By contrast, human judgment, depicted on the right, is slower but more explainable and accountable. Humans arrive at answers through layered reasoning that draws on evidence, formulates justifications, and makes inferences before reaching conclusions.

This contrast makes clear why human oversight is indispensable. Skilled users must apply domain expertise and critical skepticism to AI outputs by testing assumptions, spotting flaws, and stepping in when systems drift. Just as supervisors have always reviewed the work of junior staff, we now need a much broader cohort trained to oversee a new kind of colleague: fallible AI machines.

Human input involves refining prompts, supplying additional data, and providing independent verification. The lesson for education is unmistakable: LLMs and chatbots are no substitute for a solid foundation in domain knowledge and logical reasoning, any more than junior associate in a law firm would be a substitute for a seasoned partner. Students must be trained not only to use AI tools but also to interrogate them so they learn when to trust, when to challenge, and when to disengage altogether. Only then can AI serve as a reliable partner rather than an unaccountable oracle.

Another growing concern is how ubiquitous AI assistance may reshape human cognitive development. As shown in Figure 14, there is an "optimal zone" in which AI-augmented learning enhances engagement and productivity. In this middle range, AI supports students by accelerating routine tasks and offering guidance, while learners still grapple with core reasoning, problem-solving, and conceptual understanding.

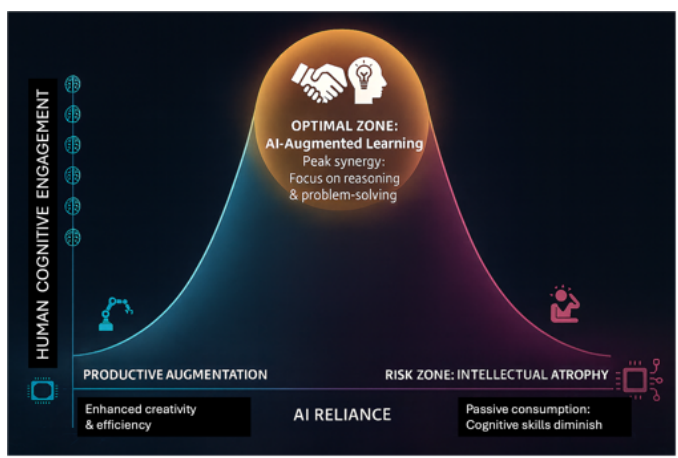

*Figure 14: The Sweet Spot of AI Reliance.*

However, Figure 14 also highlights the risks of imbalance. Too little reliance on AI may limit productive augmentation, leaving students without access to powerful tools that can deepen learning. Conversely, too much reliance on AI pushes learners into the "risk zone" of intellectual atrophy, where foundational skills erode because AI does all the cognitive heavy lifting. This "use it or lose it" dilemma raises urgent pedagogical questions: will the next generation, raised in an era of on-demand AI answers, still develop the resilience and independence needed for complex problem-solving? To safeguard against this, educators and parents must actively design learning environments where AI augments rather than replaces human thinking.

These educational challenges intersect with broader ethical and societal concerns. Preserving academic integrity becomes harder when AI can complete entire assignments with minimal oversight. Authorship and attribution grow murky when content is generated collaboratively between humans and machines (Lemley & Ouellette 2025). Bias in training data raises the risk of perpetuating harmful stereotypes or flawed conclusions, while the environmental costs of training large-scale AI models demand accountability and sustainability. Further complicating the picture are questions of misuse, ranging from disinformation campaigns to malicious automation. These issues are now at the forefront of academic and policy debates, including those underway at William & Mary (W&M 2025), as institutions grapple with how to ensure AI empowers human potential without undermining it.

Natural language programming heightens—rather than replaces—the need for human judgment. As shown in Figure 15, now that AI automates many routine CS and DS tasks, from generating boilerplate code and cleaning datasets to summarizing content and assessing programming assignments (Grandel et al. 2025), human experts are freed to concentrate on higher-level responsibilities. The real frontier isn't producing more code or faster calculations, but strengthening uniquely human capacities like error detection, ethical reasoning, requirements definition, and navigating ambiguity.

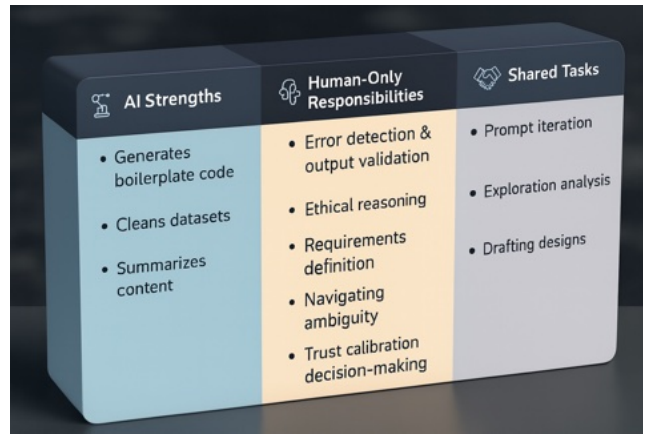

*Figure 15: AI-Human Responsibility Boundary Chart.*

Human strengths increasingly lie in higher-order judgment, complementing the rapid generative power of LLMs. AI can flood us with possibilities in seconds, but humans remain—at least for now—the best positioned to evaluate which options are valid, trustworthy, and socially responsible. As Figure 15 shows, tasks like prompt iteration, exploratory analysis, and design drafting are becoming shared efforts: AI accelerates the work, while humans provide direction and accountability. In this emerging division of labor, computational thinking isn't replaced but redefined, expanding to include systems thinking, design judgment, and ethical discernment alongside AI's raw generative capacity.

This reframing underscores why a well-rounded education matters more than ever in the age of generative AI. It is not enough for graduates to master technical tools, they must also cultivate the distinctively human faculties that AI cannot yet replicate, including ethical reasoning, the ability to justify decisions in complex and ambiguous contexts, and the skill of calibrating trust in both people and technologies. These capacities transform students from passive users of AI into active stewards of human-AI partnerships. In such collaborations, algorithms accelerate the routine and optimize efficiency, while human judgment anchors meaning, purpose, and responsibility.

When these strengths converge, the result is not simply faster computation but deeper insight and more resilient decision-making, which are collaborative outcomes that neither humans nor AI could realize on their own (Carleton et al. 2021). In this sense, computational thinking is no longer confined to abstract logic or coding. Instead, it is evolving into richer, more integrative practices that fuse technical fluency with ethical reflection, creativity, and humanistic perspective.

## 7. Concluding Remarks: Adapting to the Unstoppable Tide

The rapid spread of generative AI marks not an isolated disruption, but the latest chapter in a long history of commoditization in computing, as shown in Figure 16. Each wave has lowered barriers: spreadsheets in the 1980s gave office workers new analytic power; search engines in the 1990s and 2000s opened the world's knowledge to anyone online; low-code tools in the 2010s let non-specialists build apps with ease. Now, in the 2020s, LLMs and AI-augmented IDEs are democratizing high-level problem solving itself, enabling people across disciplines to tackle challenges once reserved for expert programmers and data scientists.

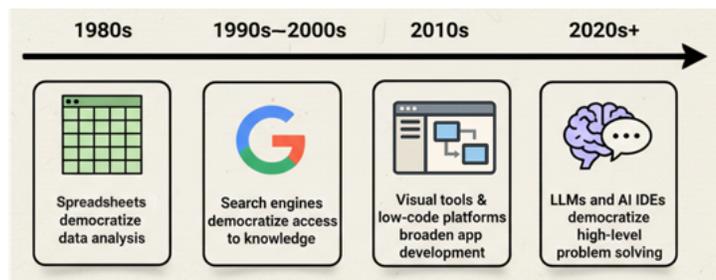

*Figure 16: Commoditization of Computing Over Time.*

This trajectory underscores a fundamental shift: generative AI is not an anomaly, but the next stage in computing's long evolution toward broader accessibility—the commoditization of computational thinking itself. Just as earlier waves of computing moved from punch cards to personal computers to smartphones, AI represents a further lowering of the barriers to entry. To resist this transformation—whether in education, government, or industry—would be akin to King Canute ordering back the tide. The tide of commoditization cannot be reversed; it can only be harnessed. The imperative, therefore, is not to preserve old gatekeeping models, but to ensure that this newfound accessibility is directed toward constructive and responsible ends.

As shown in Figure 17, commoditization works best when three domains converge: *AI capabilities* (e.g., speed, scale, generative power), *engineering discipline* (e.g., prompt design, validation, systematic practice), and *human insight* (e.g., ethics, creativity, domain knowledge). The overlap of these domains produces *amplified problem solving*, a vision of computational thinking accessible not only to expert programmers but to students, professionals, and citizens across disciplines. Reimagining curricula in this light means teaching computing as a shared language of problem solving—one that balances efficiency with responsibility and equips people everywhere to participate meaningfully in the AI era. In this way, the commoditization of computational thinking becomes both a pedagogical mandate and a societal opportunity.

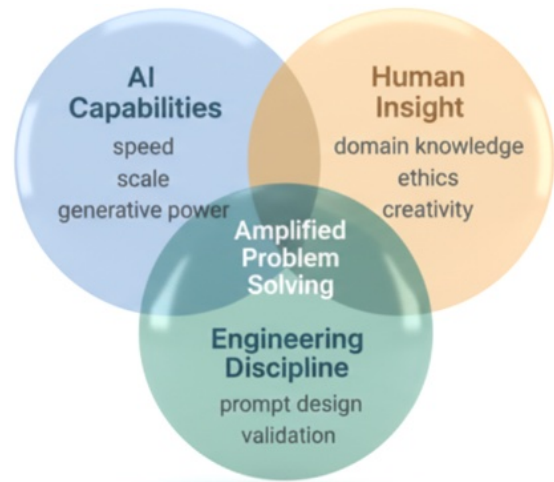

*Figure 17: What Sets AI-Augmented Humans Apart.*

The AI-augmented tide is rising; our task is to learn how to surf its turbulent waters, rather than being left behind or overwhelmed by it. By embracing this change—and guiding it responsibly—we can ensure that the next generation of computational thinkers uses AI not to replace human creativity and judgment, but to amplify them. This emerging paradigm holds the potential to commoditize computing but also raises urgent questions about the future of the computing profession and how we prepare the next generation of technologists.

**Acknowledgements**